\documentclass[doublecol]{epl2} 

\usepackage{amsmath, amssymb, graphics, graphicx}
\usepackage{textcomp, subfigure}\usepackage{color}

\definecolor{linkcolor}{rgb}{0,0,0.6} 
\usepackage[pdftex,colorlinks=true, pdfstartview=FitV, linkcolor= linkcolor, citecolor= linkcolor, urlcolor= linkcolor, hyperindex=true,hyperfigures=true]{hyperref} 

\begin{document}
\title{\bf  Dynamics of a Liquid Crystal  close to the  Fr\'eedericksz transition} 
\author{ A. Caussarieu, 
A. Petrosyan, S. Ciliberto 
\footnote{Corresponding Author:  sergio.ciliberto@ens-lyon.fr} 
}
\institute{
Universit\'e de Lyon \\
Ecole Normale Sup\'erieure de Lyon, Laboratoire de Physique ,\\
C.N.R.S. UMR5672,  \\ 46, All\'ee d'Italie, 69364 Lyon Cedex
07,  France\\}


\begin{abstract}
{{ We study experimentally and numerically the dynamics of the director of a liquid crystal driven by an electric field  close to the critical point of the Fr\'eedericksz Transition (FT). We show that the  Landau-Ginzburg (LG) equation, although it  describes correctly the stationary features of FT in a rather large range of the control parameter,  cannot be used to describe the dynamics in the same range.  }
The reasons {of this discrepancy} are related  { not only to } the  approximations done to obtain this equation {but most importantly}  to the finite value of the anchoring energy and to small asymmetries  on boundary conditions. The difference between static and dynamics is discussed.These results are useful in all of the cases where FT is used as an example for other orientational transitions. }
\end{abstract}

\maketitle

Transitions between different orientational orders appear in several systems characterized by strong anisotropy such as for example biological systems\cite{Joanny,Ramaswamy}, anisotropic phase in superfluids\cite{Fisher2007,Mullen1994}, ferromagnetic \cite{ferromagnet} and  elastic media  \cite{elastic-order}.  Liquid crystals (LC), being constituted by elongated molecules, have a strong anisotropy of their physical properties, and are  certainly the most common and general system where such a kind of transitions can be observed \cite{DeGennes,Oswald}.  For example, a nematic liquid crystal, whose molecules are initially homogeneously aligned between two parallel plates, undergoes a
transition to an elastically deformed state when a sufficiently high external electric, magnetic or optical field $E$ is appropriately applied. This is  the Fr\'eedericksz transition (FT) 
characterized by its critical field $E_c$ ; this transition is  very important,  not only for its obvious industrial applications, but also because it is used as an example to understand other systems.  
The relevant order parameter of the FT is  the unit pseudo vector $\vec{n}$ (the director) which defines the local direction of alignment of the molecules.  A stability analysis at the mean-field level of  FT shows that the  transition is of second order and that the dynamics of the order parameter can be described by a Landau-Ginzburg (LG) equation for $\vec{n}$ \cite{DeGennes,Oswald}, $E$ being the control parameter.   

The purpose of this letter  is to show (experimentally and numerically) that although the static equilibrium measurements seem to agree with the LG, the experimental study of the fluctuations and the dynamics of
 $\vec{n}$  demonstrates that such a  model {does not describe correctly the time dependent behavior}.
{ This is a useful information because, even if the purpose of the LG is to give  the threshold of the instability, 
it is often used in literature to predict the dynamics close to the critical point of the FT.}  

We consider in this letter the dynamics of  the FT of a nematic liquid crystal (NLC), subjected to an electric field $\vec{E}$~\cite{DeGennes,Oswald}, but the results are general enough to be applied to other systems where FT is used as a reference of orientational instability. In order to fix the framework of this letter, let us recall that FT must not be confused with electroconvective instabilities    { because in  FT, no stationary fluid motion exists}. However  time dependent  hydrodynamic effects, such as the backflow,  may eventually influence  the dynamics of the FT and must be taken into account. 

Because of their importance, the static properties of FT have been widely studied both theoretically\cite{Deuling} and experimentally \cite{Faetti} and the main mechanisms are well understood.  On the contrary, the study of the characteristic times of the dynamics of $\vec{n}$ above threshold, which is also very important, did not receive the same attention.
 In ref.\cite{Pieranski}, the growth rate has been measured, but, as we will see, this is a different information than the dynamics 
 {above threshold, i.e for $E\ge E_c$. } 
The properties of fluctuations above the threshold of FT have been studied through light scattering in ref.\cite{Galatola1992} but the characteristic times were not studied, and no comparison with theory has been done.  
{In ref.\cite{Zimmermann}, the relaxation time above threshold has been measured and a detailed analysis of fig.9a) of  ref.\cite{Zimmermann}  shows a discrepancy between the measured characteristic times and  those theoretically estimated. The article does not discuss this inconsistence.}
Finally Zhou and Ahlers \cite{Zhou2004} pointed out that there were problems in modeling the FT as a second order phase transition. They explained this with a random driven first order phase transition. Our observations strength the experimental observations of ref.\cite{Zhou2004} but show that the explanation is different from the one proposed in that reference. Indeed we explain the main discrepancies between theory and observations with boundary effects, which wipe out all the critical region. 
\begin{figure}[h]
 \centering
\includegraphics[width=0.8\linewidth]{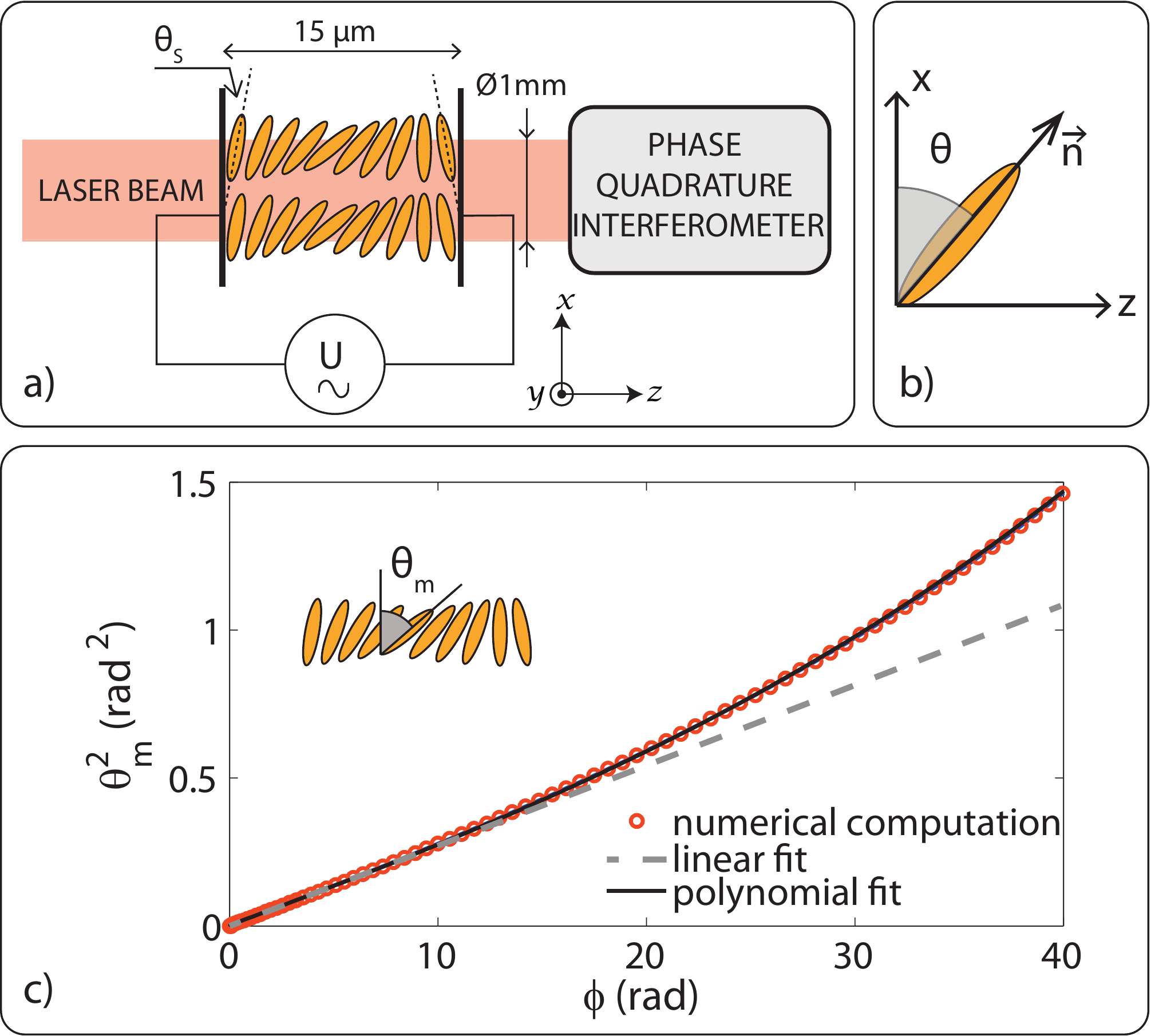}
\caption{   {a) The geometry of Fr\'eedericksz transition : director configuration for $U_0
>U_c$. Experimental setup:  a polarized  laser beam crosses  the LC cell.  The optical anisotropy of the LC  induces an  optical phase shift $\phi$ between the ordinary an extraordinary polarizations of the laser beam. A polarization interferometer measures $\phi$ (see text) ~\cite{Bellon02}. }
b) Definition of the angular displacement $\theta$ of  $\vec{n}$. c)    {Dependence of   $\theta_m^2$    {(the maximum of $\theta^2$)} on  $\phi$ used to calibrate the measure (see text), and to retrieve the valued of $\theta_m$ from the measure of $\phi$. }  } 
\label{fig1}
\end{figure}

The system under consideration is a NLC having a positive dielectric anisotropy (p-pentyl-cyanobiphenyl, 5CB, produced by Merck).  The LC is confined between two parallel glass plates at a distance $L=15\mu m$ (see fig.~\ref{fig1}). 
The surfaces of the confining plates in contact with LC have transparent Indium-Tin-Oxyde (ITO) electrodes to apply the electric field. Furthermore, to induce parallel alignment of the directors at the surfaces, a thin layer of polymer (PVA) is deposited and mechanically rubbed in one direction. Therefore, all the molecules in the vicinity of the plates have their director $\vec{n}$  parallel to the $x,z$ plane ; $\vec{n}$ can be written $\vec{n}=(\cos\theta(z),0,\sin\theta(z))$ (see fig.~\ref{fig1})~\cite{anchoring,cognard}, where $\theta(z)$ is the angle between the director and the surface. 
In the absence of any electric field, the functional form of $\theta(z)$ is determined by the boundary conditions at $z=0$ and $z=L$, that is by the pretilt angle $\theta_s$ between  the director of the molecules anchored on the surfaces and the rubbing direction (see fig.\ref{fig1}a).  For the 5CB in contact with PVA, $\theta_s\simeq 0.05\,rad$. During the assembling of our cell, the rubbing directions on the two plates  have been oriented for obtaining an antiparallel alignment \cite{yeh-book}, which imposes $\theta(L/2)=0$ :  $\theta(0)=-\theta(L)=-\theta_s$ and $\theta(z)=(2z/L-1)\theta_s$ at $E=0$. 
\footnote{This is not the most common configuration : indeed the parallel one, i.e. $\theta(0)=\theta(L)$, is the most used  because it induces a tilt in the center of the cell which facilitates the FT at a value of the  control parameter $E$ much smaller than the theoretically  predicted value \cite{meyerhofer,Stewart-Book}. In our experiment we used the antiparallel because theoretically it should give a sharp transition, as we will see in the following.}
The LC is then submitted to an electric field perpendicular to the confining  plates. To avoid the electrical polarization of the LC, the electric field has a zero mean value which  is  obtained by applying a sinusoidal voltage $V$ at a frequency of $f_d=10$~kHz between  the ITO electrodes, i.e. $V = \sqrt{2} U_0\cos(2 \pi \cdot f_d \cdot t)$~\cite{DeGennes, Oswald}.
 
With these experimental constrains, the free energy per unit surface of  the  LC   takes the form \cite{Deuling,Faetti}: 
\begin{eqnarray}
F_s &=& {k_1  \over 2} \int_o^L \left[ (1+k\sin^2(\theta(z))) \left( {d \theta(z) \over dz} \right)^2  \right] dz  + \notag  \\ 
 & -&  Ê{U_0^2  \  \epsilon_\perp \over 2  \int_o^L {dz\over 1+\varUpsilon \sin^2\theta(z)}   } 
\label{eq:free-energy} 
\end{eqnarray}
where  $k=(k_3-k_1)/k_1$ and $\varUpsilon=(\epsilon_\parallel-\epsilon_\perp)/\epsilon_\perp$ are respectively the elastic and dielectric anisotropy parameters of the LC, with  $k_i$ ($i=1,3$) its elastic constants, $\epsilon_\parallel$ the parallel dielectric constant and  $\epsilon_\perp$ the perpendicular one. 

The FT, in the vicinity of the threshold, is usually described by the LG equation obtained from equation (\ref{eq:free-energy}) \cite{DeGennes, Oswald}. 
In fact, one assumes $\theta_s=0$ and the sinusoidal form of the solution $\theta(z,t)=\theta_m(t)\sin(\pi z/L)$ ; then the free energy can be developed to fourth order in $\theta_m$.
 In this way, one gets an expression (\ref{eq:F_Landau}) for the free energy, where 
 $\varepsilon=(U_0/U_c)^2-1$ is the reduced control parameter  and 
  $U_c=\sqrt{ k_1\pi^2/(\epsilon_\perp \varUpsilon)}$ the critical voltage for FT \cite{DeGennes,Oswald}. In order to have a precise comparison we recall that the commonly accepted values for 5CB  for these parameters are : $U_c=0.710V$, $\kappa=0.36$ and $\varUpsilon=2$ and $k_1=6.15 \ 10^{-11}$N.
\begin{equation}
F_s=\frac{\pi^2k_1}{2L}\left[-\frac{\epsilon_0\epsilon_\bot U^2}{\pi^2k_1}-\frac{\theta_m^2}{2}\varepsilon+\frac{\theta_m^4}{8}\left(\kappa+1+\varUpsilon\right)\right]
\label{eq:F_Landau}
\end{equation}
The dynamical equation for $\theta(z)$  is $\gamma d\theta /dt = -\delta F_s / \delta \theta$ where $\gamma$ is the rotational viscosity of the LC \cite{DeGennes,Oswald}. Introducing the characteristic time $\tau_0=\frac{\gamma L^2}{\pi^2k_1}$, the dynamical equation of $\theta_m$ is: 
 \begin{eqnarray}
\tau_0 \frac{{\rm d} \theta_m}{{\rm d} t}=  \varepsilon \ \theta_m -
{1 \over 2} (\kappa+\varUpsilon+1) \theta_m^3 +\eta
\label{momentum_equation}
\end{eqnarray}
where $\eta$ is a thermal noise delta-correlated in time \cite{Sanmiguel1985} describing  the director thermal fluctuations. 
  Eq.\ref{momentum_equation}, whose stationary solution is $\theta_o^2=2\varepsilon/ (\kappa+\varUpsilon+1)$, shows that if $\theta_m$ remains small, then its dynamics is described by a LG equation and one expects  mean-field critical phenomena~\cite{DeGennes, Oswald, Sanmiguel1985}. 
   Indeed from eq.\ref{momentum_equation}, calling $\delta\theta$ the thermal fluctuations around  $\theta_m$, i.e.  
  {$\theta_m=\theta_0+ \delta\theta$, we can write a Langevin equation for $\delta\theta$ : 
 $\tau_0  \dot\delta\theta =-2\varepsilon \delta\theta +\eta$, which implies that the linear response time of the system}
 is $\tau=\tau_0/(2\varepsilon)$ and the variance is 
 $<\delta\theta^2>\propto k_BT/( 2\varepsilon) $,
  where $k_B$ is the Boltzman constant and $T$ the temperature. 
 However, eq.\ref{momentum_equation} is a crude approximation and in the following we want  to understand to 
 which extent, in a real system, the dynamics of $\theta(z)$ is well described by this equation. This is an important and useful question because
  the FT is used as  a model of transition between different orientational  orders.

Let us now describe how $\theta(z)$ is  measured in our experiment, which is sketched in fig.\ref{fig1}a).  
The deformation of the director field produces an  anisotropy of the refractive
index of the LC cell. 
This optical anisotropy  can be precisely estimated by measuring the optical phase shift $\phi$ between a light beam crossing the cell linearly polarized along $x$-axis (ordinary ray) and another beam crossing the cell polarized along
the $y$-axis (extraordinary ray). 
In our experiment ( fig.\ref{fig1}a) a laser beam of radius $1$mm 
produced by a stabilized He-Ne laser ($\lambda = 632.8$~nm) crosses the cell; the beam is normal to the cell and linearly polarized at 
$45^{\circ}$ from the $x$-axis.
The optical phase shift $\phi$ between the ordinary and extraordinary beams, is measured by a very sensitive polarization interferometer~\cite{Bellon02}. 
The   phase shift $\phi$ can be expressed  in terms of  the maximum $\theta_m^2$ of $\theta^2(z)$, integrating numerically 
the non-linear  equation of ref.\cite{Deuling}. The results for our experiment is plotted in fig.\ref{fig1}c). Using the above mentioned sinusoidal approximation for $\theta_z$, 
 we find $ \phi=\frac{L \pi n_e(n_e^2-n_o^2)}{2 \lambda n_o^2} \theta_m^2$ with ($n_o$, $n_e$) the two anisotropic refractive indices. Notice that the use of the interferometer allows a quantitative measure of $\theta_m^2$ 
 as a function of $\phi$ because all the other parameters are known (see ref\cite{Caussarieu} for details). This linear approximation  is compared in fig.\ref{fig1}c) with the general solution computed for the parameters of our experiment using the equations of ref.\cite{Deuling}. The linear  approximation is very good for  $\theta_m^2<0.3\, rad ^2 $. However using a polynomial fit,  the numerical solution can be reversed to compute  $\theta_m^2$ from the measure of $\phi$. 
The phase $\phi$, measured by the interferometer,  is acquired with a resolution of $24$ bits at a sampling rate of $1024$ Hz.
The instrumental noise of the apparatus~\cite{Bellon02} is  three orders of magnitude smaller than the amplitude $\delta \phi$ of the fluctuations of $\phi$  induced by the thermal fluctuations  of $\theta_m$.
   {The fact that $\phi\propto \theta_m^2$ has  important consequences 
in the measure of the thermal fluctuations of  $\theta_m$ (see ref.\cite{Caussarieu}), because  
 $\phi=\phi_o+\delta \phi \propto (\theta_o^2 + 2 \theta_o \delta \theta) $ where $\phi_o$ and $\theta_0$ are the stationary values of $\phi$ and $\theta_m$. Thus one finds that $\delta \phi$ is related to  $\delta \theta$ as:  $\delta \phi=2 \theta_0 \delta \theta $ }

\begin{figure}[htbp]%
\includegraphics[width=\columnwidth]{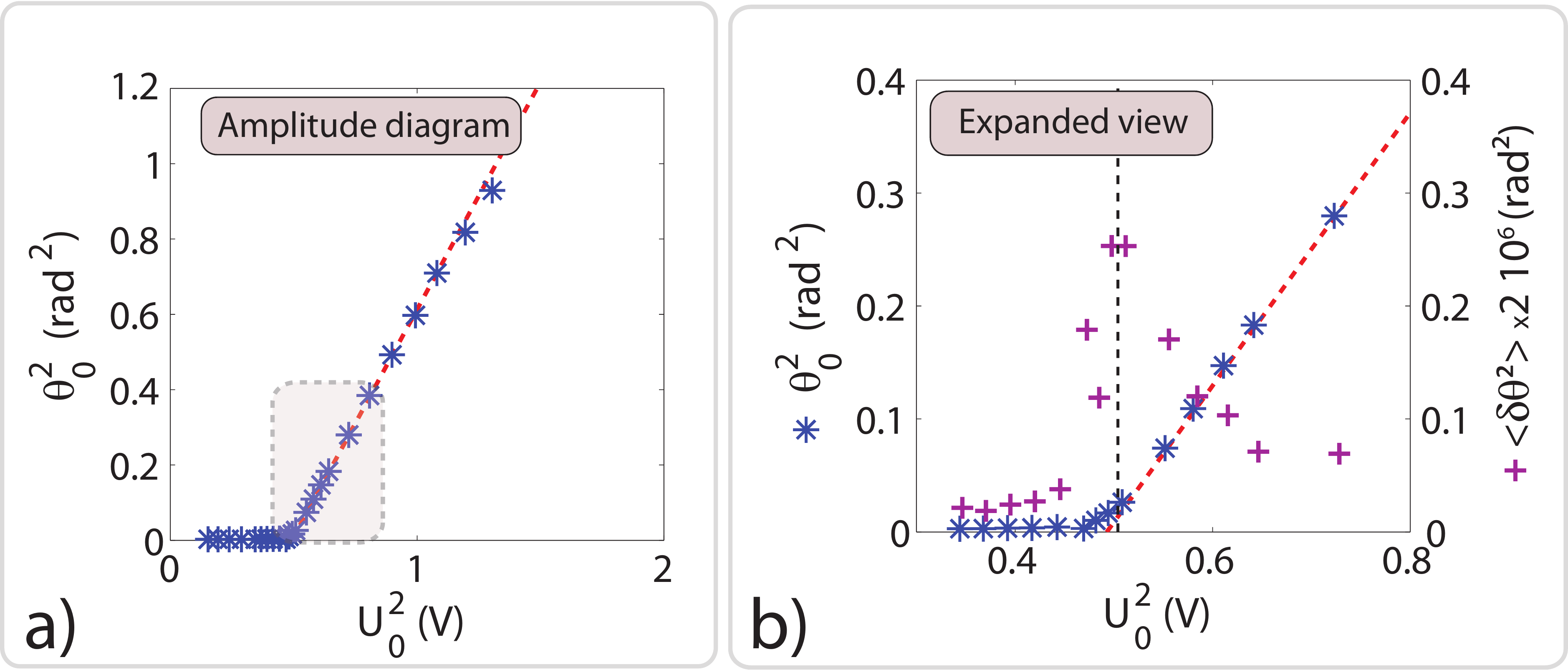}
\caption{a) Amplitude diagram of $\theta_0^2$ versus $U_o^2$. Experimental data($\star$) and solution of LG equation with $U_c=0.704V$ (red dashed line) b) Expanded view of the plot in (a). The variance $<\delta\theta^2>$ of  $\theta_m$ is plotted in b) as a function of  $U_o^2$. As the amplitude of the fluctuations of $\theta_m$ is very small the plotted values ($+$) correspond to $<\delta\theta^2>\times 2 \,10^6$.  }
\label{fig2}%
\end{figure}
Let us first discuss the experimental results shown  in fig.\ref{fig2}. In fig.\ref{fig2}a) we plot the measured $\theta_0^2$ as a function of $U_0^2$. On the same figure, the red dashed line represents the stationary solution of  eq.\ref{momentum_equation},
: this approximated solution seems to fit the data {within the interval $0.5V^2<U_0^2<1V^2$ (corresponding to $0<\varepsilon<1$), which is rather large taking into account the crude approximations done to obtain eq.\ref{momentum_equation}. 
An expanded view of fig.\ref{fig2}a) around $U_0 \simeq U_c$(i.e. $\varepsilon \approx 0$)} is plotted in fig.\ref{fig2}b) where the imperfection of the transition can be seen. The dashed red line corresponds to the best fit with the LG solution which is obtained for $U_c=0.704V$.  In that figure we also plot $<\delta\theta^2> \times 2 \, 10^6$ measured in the experiment (+).  This quantity    {does not diverge as predicted by eq.\ref{momentum_equation}. However it } has a peak exactly at $U_0=0.710V$ which defines the critical threshold
\footnote{   {This  accurate estimation of $Uc$, based on the measure of the variance  without using  any fit \cite{Caussarieu},  
agrees with previous results \cite{Faetti} }},
used to calculate $\varepsilon$ in this letter. 
It appears clearly that the the transition is not sharp and the real curve is shifted towards smaller values of $U_0$, showing that the transition occurs before the expected threshold. 

\begin{figure}[htbp]%
\begin{center}
\includegraphics[width=0.9\columnwidth]{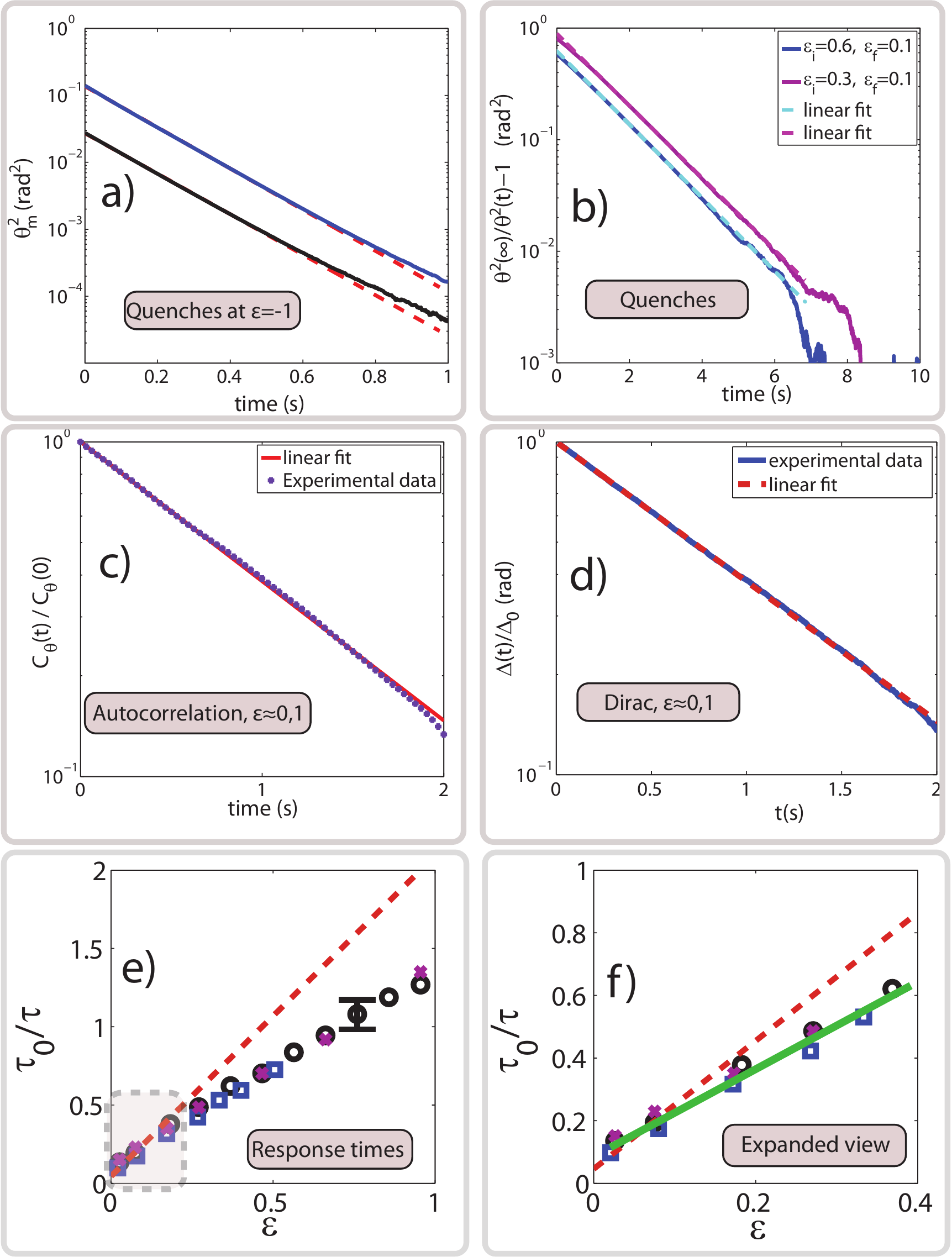}%
\end{center}
\caption{a)Evolution of $\theta_m^2$ as a function of time during quenches at zero field. Solid lines represents experimental datas whereas dotted lines represents each fit. The upper curve corresponds to $\varepsilon_i=0.2$ and the other one corresponds to  $\varepsilon_i=0.03$. The slopes of the fits are -7.1 and -7.0 s$^{-1}$ respectively. b)  Quenches in $\varepsilon$ of very small amplitude ; c) Autocorrelation function ; d) Linear response to an impulsionnal perturbation in $\varepsilon$ ; e) Normalized response time \textsl{vs} $\varepsilon$ : experimental data from quenches ($\circ$), auto-correlations ($\times$) and Dirac ($\Box$). The red dashed line corresponds to the LG prediction (eq\ref{momentum_equation}). f) Expansion of e) and linear fit of the data (continuous green line).  %
 }%
\label{fig3}%
\end{figure}

{To study the dynamics of the system, we start measuring $\tau_0$ with the standard technique \cite{Wu,Pieranski} of the quench at zero field} ($\varepsilon=-1$) starting at an $\varepsilon_{1}$ in the interval $[0 \ 0.1]$. The decay of $\theta_m^2$  at long time after the quench should go, on the basis of eq.\ref{momentum_equation}, as $\exp(-2t/\tau_0)$. The results of the measurements are reported in  fig.\ref{fig3}a) where the dependence of $\theta_m^2$ as a function of time after the quench is plotted for two different initial values of $\varepsilon_1$. 
We see that the decay rate is independent of $\varepsilon_1$ and that $\tau_0=0.28\pm0.01\,$~s. From this value and the definition of $\tau_0$ one gets that $\gamma= 0.078\pm0.005$ Pa.s, which is close to the values  reported in literature for 5CB     {  ($\gamma=0.08$ Pa.s with no  error bars \cite{Wu}). Notice that the high resolution of our measurement allows the estimation of  well defined error bars on the value of $\gamma$. }

   {As $\tau_o$ is  known, we can focus on  the dependence  of the characteristic time  $\tau$ on $\varepsilon$ for $\varepsilon>0$. 
In order to be sure of the estimated values  of $\tau$,} we measure it through three different quantities : 1) The decay rate after a quench in $\varepsilon$ of very small amplitude; 2) The characteristic time of the autocorrelation function of the thermal fluctuations of $\theta_m$ ; 3) The linear response to a Dirac perturbation of $\varepsilon$. 
  The dependence of $\theta_m^2$ as a function of time for a quench of $\delta \varepsilon=0.01$ starting at two different initial values $\varepsilon_1$ are plotted in fig.\ref{fig3}b). From the long time behavior one gets the $\tau$ at $\varepsilon_1-\delta\varepsilon$. 
  These values are plotted in figs.\ref{fig3}e-f) as a function of $\varepsilon$. In figs.\ref{fig3}c-d) we also show that the autocorrelation function  and the response to a perturbation relax with the same characteristic time when taken at the same $\varepsilon$. Repeating the measure for different $\varepsilon$, one can get the evolution of these characteristic times as a function of $\varepsilon$. The results normalized by $\tau_0$ are plotted on fig.\ref{fig3}e) and we can clearly see that the measured values of $\tau$ are independent on the method as it is enhanced by the continuous line on fig.\ref{fig3}f). 


  In  figs.\ref{fig3}e-f) the prediction of the eq.\ref{momentum_equation}, i.e. $\tau_0/\tau=2 \varepsilon$, is also plotted for comparison (red dash line). We clearly see that even for $\varepsilon<1$ , where eq.\ref{momentum_equation} seems to reproduce the data of $\theta_m$ (see fig.\ref{fig2}a), the measured $\tau_0/\tau$ are about thirty percent smaller than the prediction. Furthermore $\tau_0/\tau$ does not vanish when $\varepsilon=0$ as predicted by eq.\ref{momentum_equation} .

{To summarize the experimental data in the dynamical regime, we see that two points cannot be explained by the LG equation even for small values of $\varepsilon$ : 1) the non divergence of the response time; 2) the deceleration of the system for $\varepsilon<1$ ($\tau_0/\tau$ smaller than what was expected from LG). 
{To understand these facts, one cannot neglect the role played by the boundary effects and the non linearities during the dynamics. Therefore,} we need to write realistic boundary condition that take into account both the anchoring surface energy  $W$ and the pretilt angle $\theta_s$. For small $\theta_s$, the boundary conditions for the torque \cite{Faetti,Rapini}} are: 
\begin{eqnarray}
\left[k_1(1+k\sin^2\theta(z))  {d \theta(z) \over dz} +W(-\theta(z)\pm \alpha(z)\theta_s)\right]_{z=0,L}&=&0 \notag \\
& &\label{eq:boundarycond} 
\end{eqnarray}
where the $\pm$ correspond to $z=0$ and $z=L$ respectively (antiparallel alignment). { The parameters $\alpha$ allow us to take into account small experimental alignment defects, produced during the assembly of the cell, which  make $\theta(0)\neq-\theta(L)$. In the case of an ideal alignment, we have $\alpha(0)=\alpha(L)=1$, otherwise the ratio $SR=\alpha(0)/\alpha(L)$ is different from 1 and $SR$ identifies the magnitude of the alignment defect, i.e. $\theta(0)=-SR   \cdot  \theta(L)$.
We will show that these \textsl{asymmetries} play a crucial role in the dynamics and the static of $\theta$ close to the FT threshold.

The dynamical equation for $\theta(z)$  is $\gamma d\theta /dt = -\delta F_s / \delta \theta$ where $\gamma$ is the rotational viscosity of the LC \cite{DeGennes,Oswald}. From eq.\ref{eq:free-energy} one gets: 
\begin{eqnarray}
	\tau_o {d \theta \over dt} = {(\varepsilon+1) \sin (2 \theta) \over  2 \left[ \left( {1\over L}\int_0^L {dz \over (1+\varUpsilon   \sin^2(\theta) \, )} \right) (1+\varUpsilon \sin^2( \theta)) \right]^2 }  +  Ê \notag \\ \left(L^2 \over \pi^2\right) 
\left[ {\partial^2 \theta \over \partial z^2} (1+k \  \sin^2(\theta)) + {k \over 2} \sin(2 \theta)  \left({\partial \theta \over \partial z}\right)^2 \right]
 \label{eq:LC_z} 
\end{eqnarray}
For the 5CB in contact with the PVA, the real boundary conditions are  approximately $\theta_s \simeq 0.05\,$ rad and $W \simeq 3 \ 10^{-4} J/m^2$. 
In the very specific case in  which $\theta_s=0$, $W\rightarrow \infty$,  eq.\ref{eq:LC_z}  at $\varepsilon\simeq0$ becomes  the previously defined eq.(\ref{momentum_equation}).

We discuss first the {influence of the boundary conditions and non-linearities on the }stationary case; in fact, to obtain eq.(\ref{eq:F_Landau}) and eq.(\ref{momentum_equation}) we neglected their influence. }
To understand the role of $W$, $\theta_s$ and $SR$ (see eq.\ref{eq:boundarycond} ) on the transition, we perform several numerical simulations of eq.\ref{eq:LC_z}  with different boundaries conditions.  The stationary solutions $\theta_0$ of eq.\ref{eq:LC_z} are compared to the experimental data in fig.\ref{fig4}a). In the inset of fig.\ref{fig4}a) we see that the numerical solution (solid orange line) with ideal boundary conditions ($\theta_s=0$, $W\rightarrow \infty$)  fits the data in the whole interval of $\varepsilon$ \footnote{The accuracy of the numerical simulation has been checked with the direct numerical  minimization of eq.\ref{eq:free-energy} as done using  refs. \cite{Deuling,Faetti}}. We find that, for all boundary conditions,  we reproduce the data for large $\varepsilon$, their influence being strong in the vicinity of the threshold. In particular  in fig.\ref{fig4}a) we plot the stationary solution for  $SR=1$, $\theta_s=0.05\,$ rad and $W=3 \ 10^{-4} J/m^2$ (black $\lozenge$). 
{We see that the finite anchoring energy is responsible for a shift of the critical threshold but this shift is too small with respect to the experimental measured values. 
Moreover, the finite anchoring energy with antiparallel symmetric boundary conditions does not explain the  roundness of the transition : indeed the numerical data show that the transition remains sharp. }
In order to reproduce the imperfect bifurcation, observed in the experiment, {one has to introduce an asymmetry in the  boundary conditions on the two plates. Therefore, by keeping the same values of $\theta_s$ and $W$,  we fix the {\it ''asymmetry ''} at $SR=1.1$ which is a rather reasonable value. The stationary solution 
of eq.\ref{eq:LC_z} with these {\it ''asymmetric''}  boundary conditions is plotted in fig.\ref{fig4}a) (green $\circ$). It fits quite well the experimental data, indicating that our assumptions are able to reproduce the stationary behavior of the order parameter. 

   {Now, we want  to see whether these statements on the boundary conditions  are also able to explain the behavior  of   $<\delta\theta^2>$   at $\varepsilon\simeq 0$, plotted  in fig.\ref{fig2} and fig.\ref{fig4}b).  Therefore we compute  the numerical solution of eq.\ref{eq:LC_z} in  which we added a noise delta correlated both in space and time. The computed variances are plotted in fig.\ref{fig4}b) for different values of the boundary conditions. We see that the numerical solution  with  {\it ''asymmetric''} boundary conditions, i.e. $SR=1.1$ (green $\circ$), fits the data quite well whereas the   {\it ''symmetric''} one with $SR=1$ (black $\lozenge$) presents a true divergence of the variance at the critical point }
\footnote{ We do not plot directly the variance  of $\phi$, as it is usually done in literature.}.
Summarizing the stationary results, we see that the smoothness of the transition  around $\varepsilon\simeq 0$,  can be  simply explained by a small asymmetry on the anti-parallel boundary conditions which also reproduce the experimental values of the variance as a function of $\varepsilon$. 

\begin{figure}[htbp]%
\begin{center}
\includegraphics[width=.8\columnwidth]{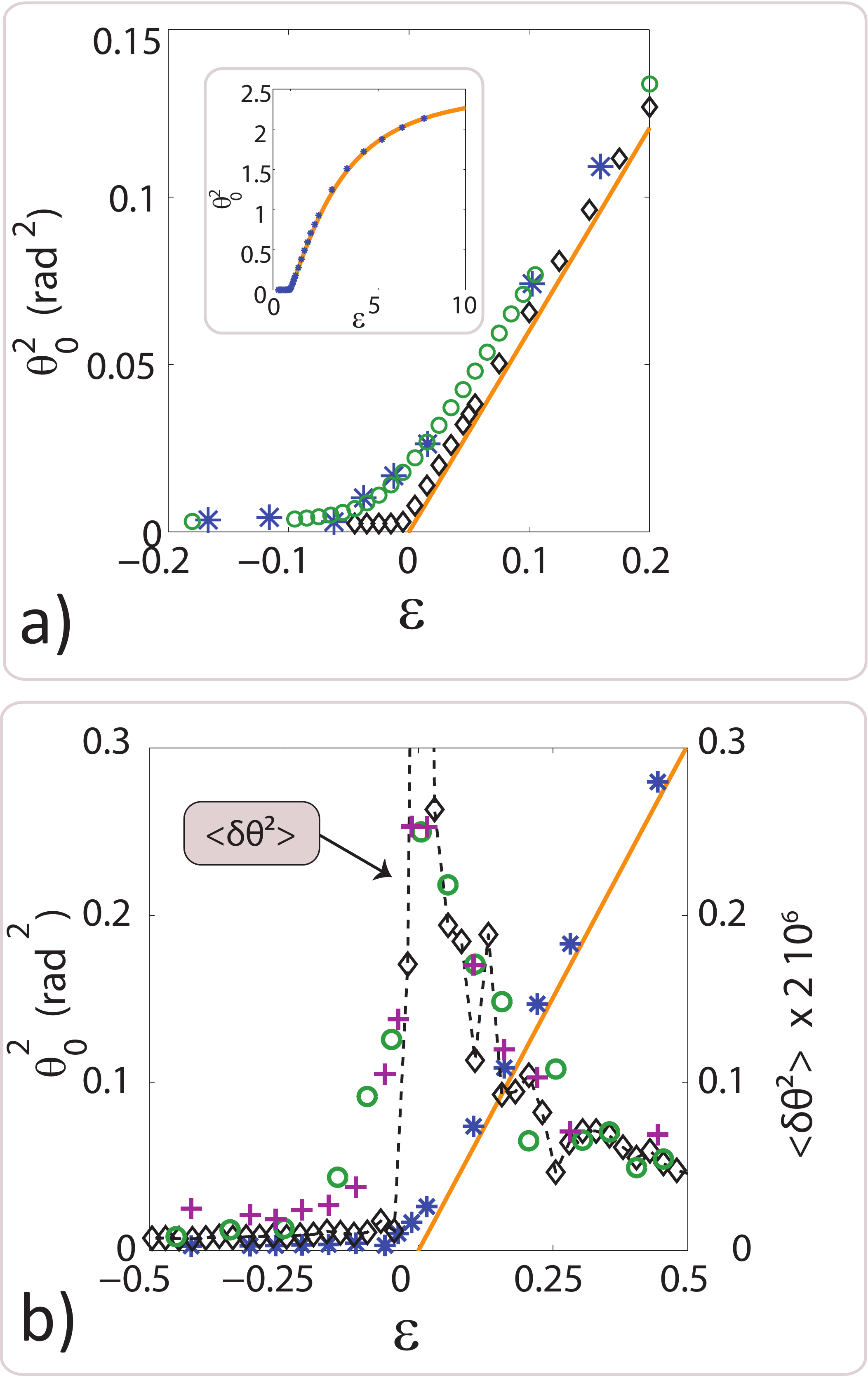}%
\end{center}
\caption{a)    {Main figure and inset : experimental $\theta_0^2$ ($*$ blue) and numerical solution of eq.\ref{eq:LC_z},  (orange solid line) with ideal boundary conditions ($\theta_s=0$, $W\rightarrow \infty$) . Numerical solution with antiparallel boundary conditions and realistic anchoring energy (SR=1, black $\lozenge$) and the same numerical solution (antiparallel) with an asymmetry of about 10 percent (SR=1.1, green $\circ$). 
b) Experimental $\theta_0^2$ ($*$ blue). The  experimental variance ($\sigma_\theta^2\times2.10^{6}$, purple $+$), is compared to the variance  (SR=1, black $-\lozenge-$ ; SR=1.1, green $\circ$) of the numerical  solution of eq.\ref{eq:LC_z} }
}%
\label{fig4}%
\end{figure}
 
We discuss now the dynamics of the system through its characteristic time.  
The question is how to explain the discrepancy between the prediction of the eq.\ref{momentum_equation} (linear dependance, dashed red line on the figure \ref{fig3}e-f) and the measured values. In order to answer to this question, we numerically integrate eq.\ref{eq:LC_z}. Making small quenches in $\varepsilon$ we measure the relaxation time of $\theta_m^2$ of this equation using the boundary conditions of $W$ and $\theta_s$ used for the symmetric and the asymmetric cases in fig.\ref{fig4}. The computed values of $\tau_o/\tau$ are  plotted in fig.\ref{fig5}. 
 We immediately see that for $\varepsilon >  0.15$ the two solutions give the same results whereas for $\varepsilon<0.1$ the asymmetric solution $SR=1.1$ fits the data confirming our hypothesis of imperfect boundary conditions. Instead the symmetric case perfectly agrees with the LG solution for $\varepsilon<0.1$.  
  This is an important statement because it means that although the solution of eq.\ref{momentum_equation} reproduces the static behavior of 
$\theta_m^2$ for $\varepsilon<1$, this equation is unable to reproduce the dynamical features in the same region.
We therefore wonder about the experimental results on the growth-rate starting from $\varepsilon=-1$ presented in ref.\cite{Pieranski}, which show the agreement with LG predictions. To check this point we performed the numerical simulation on the growth rates as done in ref.\cite{Pieranski} and we  find that the  results are exactly what LG predicts. This is due to the fact that when the instability starts, the mean value of $\theta_m$ is very small and the non-linear terms are negligible). Instead when studying the dynamics above threshold { for 
$\varepsilon>0.1$, the non-linear terms,} although negligible for the static, play an important role and they completely modify the dynamics of eq.\ref{momentum_equation}. In  the region at small $\varepsilon<0.1$, where  
the dynamics of the ideal symmetric solution agrees with that of eq.\ref{momentum_equation} (see fig.\ref{fig5}),    the experimental imperfections wipe out the LG dynamics.
Therefore one concludes that eq.\ref{momentum_equation} can never be used to have a quantitative behavior of the relaxation time in the region where the static solution seems to fit the static experimental data.
 Here we have shown only the results for anti-parallel anchoring, but in the case of parallel anchoring the roundness of the transition is larger and the effect of the dynamics induced by the imperfect bifurcation is  more important than in our case . 

\begin{figure}[htbp]%
\begin{center}
\includegraphics[width=.7\columnwidth]{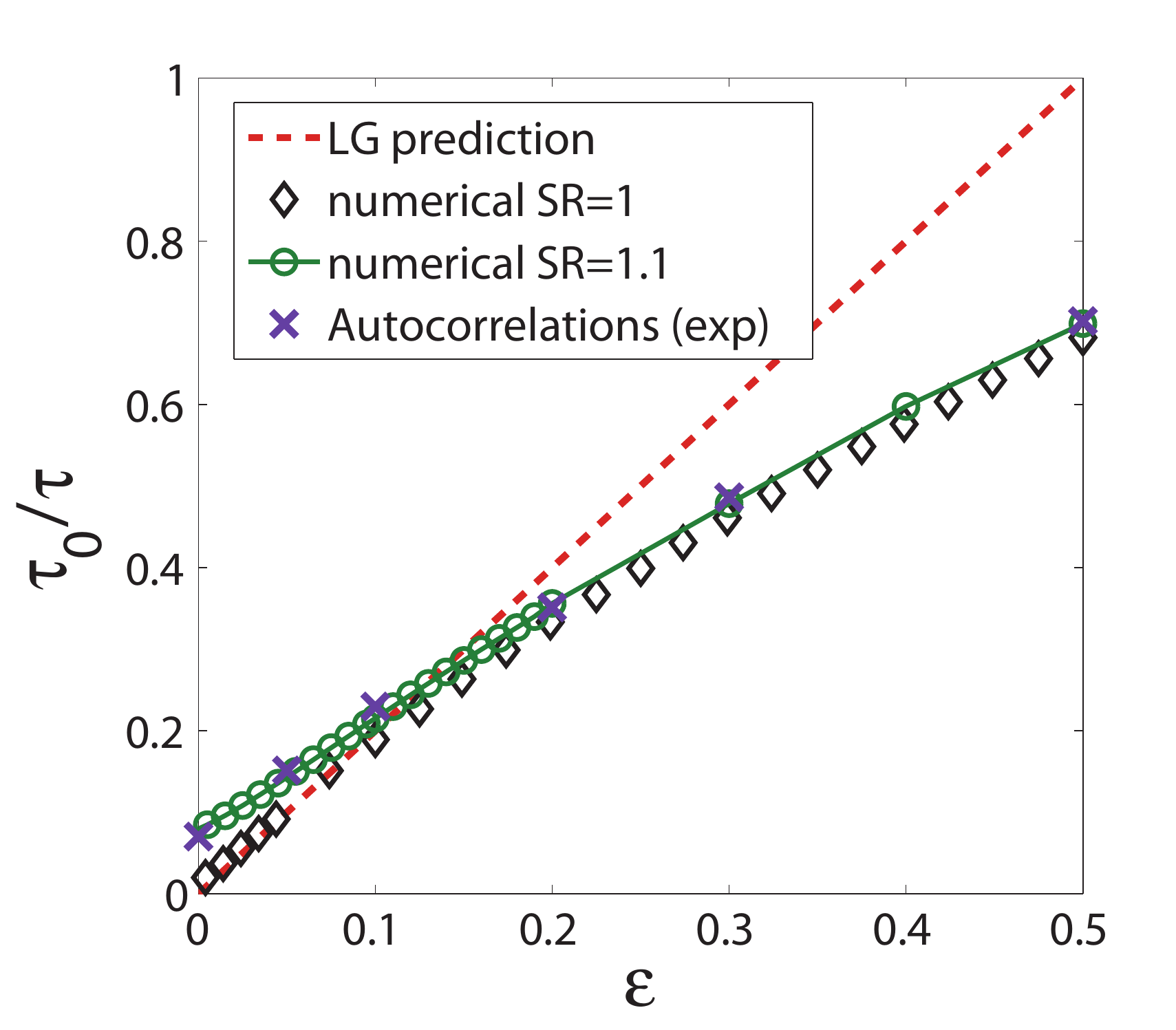}%
\end{center}
\caption{Normalized response time ($\tau_0/\tau$) as a function of  $\varepsilon$ : LG prediction (dashed red line) ; experimental data from correlations (purple $\times$). Numerical solution of eq.\ref{eq:LC_z}, (SR=1 black $\lozenge$; SR=1.1 green $\circ$ )}%
\label{fig5}%
\end{figure}

In ref\cite{Zimmermann,Brown} there are experimental studies of the dynamic of the director when the electrical field is abruptly changed above threshold. In both studies, in agreement with our observations,  the experimental data cannot be reproduced by the analytical solution of LG.  This is obvious in the light of fig.\ref{fig5} where we show that there is no region in $\varepsilon$ where the LG equation can be used to study the experimental dynamics. 

Before concluding a few words about the back-flow. For the parameters of 5CB the effect of the back flow on the dynamics is
 certainly negligible for $\varepsilon< 2$ \cite{DeGennes,Oswald,Pieranski}.  Furthermore the back-flow corresponds to an acceleration of the dynamics and not to a slower dynamics with respect to that predicted by the simple LG equation.  This is confirmed by the excellent agreement 
 between the experimental results and the numerical solution of eq.\ref{eq:LC_z}, which does not take into  account the back-flow.
 
The main conclusion of this paper is that although the LG equation has been used since several decades to study the LC dynamics close to the FT, it is actually useless, because in the region where it is valid the critical behavior is completely destroyed by small asymmetries in boundary conditions and the presence of a finite $W$. The fact that the stationary solution seems to be correct till $\varepsilon<1$ is just accidental and it is actually the origin of this misunderstanding. 
The non-divergence at the critical points had originated in the past several doubts on the nature of the FT, leading to rather complex explanation. Indeed,   it is only related to the relationship between $\delta\theta$ and the real measured variable $\delta\phi=2 \delta\theta \ \theta_0$, therefore $<\delta\theta^2>= <\delta\phi^2> /(4 \theta_0^2)$. 
As LG is used in many other fields this example is very useful in general because it 
shows that the agreement of the stationary solution does not guarantee that the equations describe correctly the dynamical behavior. 

{This work is supported by the ERC grant Outeflucop. We acknowledge useful discussion with P. Holdsworth, P. Manneville and P. Oswald}

%
%
%
%
%
%
%


\begin{thebibliography}{0}
\bibitem{anchoring} Several kinds of coating, which are known to give  strong planar alignment,  have been checked. The results are independent on the type of coating used.
\end{thebibliography}
\end{document}